\documentclass[preprint,showpacs,showkewords,preprintnumber,amsmath,amssymb]{revtex4}
 \usepackage{tipa}
 \usepackage{amsmath}
 \usepackage{txfonts}
 \usepackage{amssymb}
 \usepackage{graphicx,epsfig}
 \usepackage{bm}
 \begin{document}

 \title{ Lepton mixing patterns from the group $\Sigma(36\times3)$ with a generalized CP transformation}
 \author{Shu-jun Rong}\email{rongshj@snut.edu.cn}

 \affiliation{Department of Physics, Shaanxi University of Technology, Hanzhong, Shaanxi 723000, China}

 \begin{abstract}
The group $\Sigma(36\times3)$ with the generalized CP transformation is introduced to predict the mixing pattern of leptons.
Various combinations of abelian residual flavor symmetries with CP transformations are surveyed. Six mixing patterns could accommodate the fit data of neutrinos oscillation at $3\sigma$ level. Among them, two patterns predict the nontrivial Dirac CP phase, around $\pm 57^{\circ}$ or $\pm 123^{\circ}$,
which is in accordance with the result of the literature and the recent fit data. Furthermore, one pattern could satisfy the experimental constraints at $1\sigma$ level.
\end{abstract}

 \pacs{14.60.Pq,14.60.St}

 \maketitle

 \section{Introduction}
The oscillation experiments of reactor neutrinos~\cite{1,2,3,4,5} have determined the nonzero $\theta_{13}$, which opened the winder to explore some unknown mixing parameters including the mass ordering of neutrinos, the octant of $\theta_{23}$, and the Dirac CP phase. Among these unknowns, the nontrivial Dirac phase could provide the origin of CP violation in the lepton sector which is important for the interpretation of some fundamental questions in particle physics and cosmology such as the asymmetry of matter and antimatter. Although some fit data~\cite{6,7} hints that the Dirac phase may be maximal, i.e., $\sin\delta=-1$, the decisive evidence is still needed. In the theoretical aspect, how to predict the CP phase is interesting. Various phenomenological schemes are proposed. One of the  most popular approaches is resorting to flavor groups especially the discrete ones, see Refs.~\cite{8,9,10,11,12,13,14} and reviews~\cite{15,16} for example. Following this approach, a general flavor group $G_{f}$ is assumed for the Lagrangian of the theory. Because of the nontrivial vacuum expectation values of scalars, the original flavor symmetry is broken, namely $G_{f}$ broken to $G_{e}$ in the charged lepton sector and $G_{\nu}$ in the neutrino sector respectively. In the direct method, the mixing matrix is completely determined by the residual flavor symmetries of leptons. However, recent systematical researches~\cite{17,18} reveal that the mixing pattern determined fully by the residual flavor groups usually predicts the trivial Dirac phase. So other strategies should be considered in the construction of the leptonic mixing model. As a useful scheme, generalised CP transformations (GCP)~\cite{19,20,21,22,23,24,25,26,27,28,29,30,31,32,33,34,35,36,37,38} are introduced to extract information on CP phases. Specially, in the so called semidirect method~\cite{29,37}, the original group is $G_{f}\rtimes H_{CP}$, where $H_{CP}$ is the group of GCP acting on the flavor space. The residual symmetry in the neutrino sector is  $Z_{2}\times H^{\nu}_{CP}$. Because of the degeneracy of the eigenvalues of $Z_{2}$~\cite{39,40,41}, the mixing matrix of leptons is not completely determined. There is a real parameter to coordinate the matrix, which releases the tension between the flavor group and the fit data of neutrinos.

 In this paper, we choose the group $\Sigma(36\times3)$ with GCP to predict the lepton mixing pattern. The mixing pattern from this group has been studied in Ref.~\cite{42}. However, without corrections, it cannot commodate the fit data at $3\sigma$ level in the direct method. We find that when the GCP is considered in the semidirect method, viable mixing angles and the nontrivial Dirac CP phase could be obtained. Since the Klein group $K_{4}$ is not a subgroup of $\Sigma(36\times3)$, the cyclic group $Z_{n}$ ($n\geq2$) and $Z_{2}$ in the 3-dimensional representation are considered as the residual flavor symmetry of the charged lepton sector and that of the neutrino sector respectively.  We survey various combinations of $(Z_{n}, Z_{2})$ with the GCP, and find that two type of combinations (up to equivalent ones) are viable at  $3\sigma$ level of the constraints. There are two free parameters in the mixing matrix of leptons. Numerical analysis of the parameters show that the nontrivial Dirac CP phase is around $\pm57^{\circ}$ or $\pm 123^{\circ}$ in a type of combinations. This result is in accordance with that obtained in Ref.~\cite{41} and the fit data~\cite{46}. The predicted range of the mixing parameters is small. These results are easy to examine by the future neutrinos oscillation experiments.

 The outline of the article is as follows. In section
\uppercase\expandafter{\romannumeral2}, we summarise the basic facts of the group $\Sigma(36\times3)$ and GCP.
 In section \uppercase\expandafter{\romannumeral3}, we survey combinations of residual flavor symmetries with the GCP and give the analytical expressions for mixing angles and the CP invariants of the viable mixing patterns. Numerical results are also shown in this section.
 Finally, we present a summary.

 \section{Framework}
In this section, we recapitulate the basic facts of the group $\Sigma(36\times3)$ and summarize the approach of deriving the lepton mixing pattern on the basis of the residual flavor symmetries with the GCP in the semidirect method. Our derivations are based on the Majorana mass matrix.

\subsection{Group theory of $\Sigma(36\times3)$}
The basic facts of the group $\Sigma(36\times3)$ are taken from Refs.~\cite{42,43}.
This group has 108 elements. They could be expressed with the
generators a, b, c, which satisfy the following relations~\cite{42,43}:
\begin{equation}
\label{eq:1}
a^{3}=c^{3}=b^{4}=I,~~ab^{-1}cb=abc^{-1}b^{-1}=(ac)^{3}=I,
\end{equation}
where $I$ denotes the identity element.
The group has 14 irreducible representations~\cite{42,43}:
\begin{equation}
\mathbf{1}^{(0)},~~\mathbf{1}^{(1)},~~\mathbf{1}^{(2)},~~\mathbf{1}^{(3)},~~\mathbf{3}^{(0)},~~\mathbf{3}^{(1)},~~\mathbf{3}^{(2)},~~\mathbf{3}^{(3)},
~~(\mathbf{3}^{(0)})^{*},~~(\mathbf{3}^{(1)})^{*},~~(\mathbf{3}^{(2)})^{*},~~(\mathbf{3}^{(3)})^{*},~~\mathbf{4},~~\mathbf{4}^{\prime},
\end{equation}
where the singlets $\mathbf{1}^{(1)}$  and $\mathbf{1}^{(3)}$ are complex conjugated to each other.

In the 3-dimensional representation $\mathbf{3}^{(p)}$, the generators could be expressed as~\cite{43}
\begin{equation}
\label{eq:3}
\rho(a)=\left(
  \begin{array}{ccc}
   0 & 1 & 0 \\
    0 & 0 & 1 \\
    1 & 0 & 0 \\
  \end{array}
\right),~~\rho(b)=\frac{i^{p}}{\sqrt{3}i}
\left(
  \begin{array}{ccc}
   1 & 1 & 1 \\
    1 & \omega & \omega^{2} \\
    1 & \omega^{2} & \omega \\
  \end{array}
\right),~~\rho(c)=\left(
  \begin{array}{ccc}
   1 & 0 & 0\\
    0 & \omega &0 \\
    0 & 0 & \omega^{2} \\
  \end{array}
\right),
\end{equation}
with $p=0, 1, 2, 3$,~$\omega=e^{i2\pi/3}$.
Note that we choose the representation where $p=0$ in the following sections. The final lepton mixing matrix is not dependent on this special representation except a trivial global phase.

Since the generators of the group in the 3-dimensional representation satisfy the reordering relations~\cite{43}
\begin{equation}
\rho(ac)=\omega \rho(ca), ~\rho(a^{2}c)=\omega^{2} \rho(ca^{2}), ~\rho(ac^{2})=\omega^{2}\rho(c^{2}a), ~\rho(a^{2}c^{2})=\omega \rho(c^{2}a^{2}),
\end{equation}
and the relations~\cite{43}
\begin{equation}
\rho(bc)=\rho(ab), ~ \rho(ba)=\rho(c^{2}b),
\end{equation}
every element of the group in the 3-dimensional representation  could be expressed as~\cite{43}
\begin{equation}
\label{eq:2}
\rho(g)=\omega^{\alpha}\rho(c^{\beta})\rho(a^{\gamma})\rho(b^{\lambda}),~~with~~\alpha, \beta, \gamma=0, 1, 2,~~ \lambda=0, 1, 2, 3.
\end{equation}
Accordingly, the 14 conjugacy classes in the 3-dimensional representation are listed as follows~\cite{43}:
\begin{equation}
1C^{1}_{1}=\{~E~\},~~1C^{3}_{2}=\{~\omega E~\},~~1C^{3}_{3}=\{~\omega^{2}E~\},
\end{equation}
\begin{equation}
12C^{3}_{4}=\{~\rho(c), \omega \rho(c), \omega^{2}\rho(c), \rho(c^{2}), \omega \rho(c^{2}), \omega^{2}\rho(c^{2}), \rho(a), \omega \rho(a), \omega^{2}\rho(a), \rho(a^{2}), \omega \rho(a^{2}),\omega^{2} \rho(a^{2})~ \},
\end{equation}
\begin{equation}
\begin{array}{c}
  12C^{3}_{5} =
\{
\rho(ca), \omega \rho(ca), \omega^{2}\rho(ca), \rho(c^{2}a), \omega \rho(c^{2}a), \omega^{2}\rho(c^{2}a),
\rho(ca^{2}), \omega \rho(ca^{2}),  \\
 \omega^{2}\rho(ca^{2}), \rho(c^{2}a^{2}), \omega \rho(c^{2}a^{2}), \omega^{2} \rho(c^{2}a^{2})\},
\end{array}
\end{equation}
\begin{equation}
9C^{2}_{6} =\{~
\rho(b^{2}), \rho(ab^{2}), \rho(a^{2}b^{2}), \rho(cb^{2}), \rho(c^{2}b^{2}), \omega \rho(ca^{2}b^{2}), \omega \rho(c^{2}ab^{2}), \omega^{2}\rho(cab^{2}), \omega^{2}\rho(c^{2}a^{2}b^{2})~\},
\end{equation}
\begin{equation}
9C^{6}_{7} =\omega C^{2}_{6},~~ 9C^{6}_{8} = \omega^{2}C^{2}_{6},
\end{equation}
\begin{equation}
9C^{4}_{9} =\{
~\rho(b), \omega \rho(cb),  \omega \rho(c^{2}b), \omega \rho(ab), \omega \rho(a^{2}b), \rho(c^{2}ab), \rho(ca^{2}b), \omega \rho(cab), \omega \rho(c^{2}a^{2}b)~\},
\end{equation}
\begin{equation}
9C^{12}_{10} =\omega C^{4}_{9},~~ 9C^{12}_{11} = \omega^{2}C^{4}_{9},
\end{equation}
\begin{equation}
9C^{4}_{12} =\{~
\rho(b^{3}), \omega^{2}\rho(cb^{3}), \omega^{2}\rho(c^{2}b^{3}), \omega^{2}\rho(ab^{3}), \omega^{2}\rho(a^{2}b^{3}), \omega^{2}\rho(c^{2}ab^{3}), \omega^{2}\rho(ca^{2}b^{3}), \rho(cab^{3}),
\rho(c^{2}a^{2}b ^{3})~\},
\end{equation}
\begin{equation}
9C^{12}_{13} = \omega C^{4}_{12},~~ 9C^{12}_{14} = \omega^{2}C^{4}_{12},
\end{equation}
where the notation $iC^{j}_{k}$ denotes that the k-th conjugacy class contains $i$ elements of order $j$.

On the base of the conjugacy classes, we could obtain the information on the automorphism of the group. The structure of the automorphism group of $\Sigma(36\times3)$ is listed as follows:
\begin{equation}
\label{eq:10}
\begin{array}{c}
\mathrm{Z}(\Sigma(36\times3))=Z_{3}, ~~\mathrm{Aut}(\Sigma(36\times3))\cong (Z_{3} \times Z_{3}) \rtimes Q_{8})\rtimes Z_{2}\cong\Sigma(36)\rtimes K_{4},\\
\mathrm{Inn}(\Sigma(36\times3))\cong(Z_{3} \times Z_{3})\rtimes Z_{4}\cong\Sigma(36),~~\mathrm{Out}(\Sigma(36\times3))\cong K_{4} = \{~id, u_{1}, u_{2}, u_{1}u_{2}~\},
\end{array}
\end{equation}
where Z, Aut, Inn, Out denote the centre, the automorphism group, the inner automorphism and the outer automorphism group of $\Sigma(36\times3)$ respectively, $\Sigma(36)\equiv\Sigma(36\times3)/Z_{3}$. An inner automorphism corresponds to the group conjugation which leaves the conjugacy classes of the group invariant. In contrast, an outer automorphism swaps conjugacy classes and representations to keep the character table of the flavor group invariant~\cite{23}. In detail, the generator $u_{1}$ exchanges the conjugacy classes and the representations of the group $\Sigma(36\times3)$ as~\cite{42}
\begin{equation}
1C_{2}^{3}\stackrel{u_{1}}{\longleftrightarrow} 1C_{3}^{3},~~9C_{7}^{6}\stackrel{u_{1}}{\longleftrightarrow} 9C_{8}^{6},~~9C_{10}^{12}\stackrel{u_{1}}{\longleftrightarrow} 9C_{11}^{12},
~~9C_{13}^{12}\stackrel{u_{1}}{\longleftrightarrow} 9C_{14}^{12},
\end{equation}
\begin{equation}
~\mathbf{3}^{(0)}\stackrel{u_{1}}{\longleftrightarrow} (\mathbf{3}^{(0)})^{*},~~\mathbf{3}^{(2)}\stackrel{u_{1}}{\longleftrightarrow} (\mathbf{3}^{(2)})^{*},~~\mathbf{3}^{(1)}\stackrel{u_{1}}{\longleftrightarrow} (\mathbf{3}^{(3)})^{*},~~\mathbf{3}^{(3)}\stackrel{u_{1}}{\longleftrightarrow} (\mathbf{3}^{(1)})^{*}.
\end{equation}
Note that other conjugacy classes and representations are invariant under the action of $u_{1}$.
These transformations could be realised through the mappings as follows~\cite{42}
\begin{equation}
~a\stackrel{u_{1}}{\longrightarrow} c^{2}a,~~b\stackrel{u_{1}}{\longrightarrow} acb,~~c\stackrel{u_{1}}{\longrightarrow} ca.
\end{equation}
And the actions of the generator $u_{2}$ could be expressed as~\cite{42}
\begin{equation}
9C_{9}^{4}\stackrel{u_{2}}{\longleftrightarrow} 9C_{12}^{4},~~9C_{10}^{12}\stackrel{u_{2}}{\longleftrightarrow} 9C_{13}^{12},~~9C_{11}^{12}\stackrel{u_{2}}{\longleftrightarrow} 9C_{14}^{12},
\end{equation}
\begin{equation}
\mathbf{1}^{(1)}\stackrel{u_{2}}{\longleftrightarrow} \mathbf{1}^{(3)},~~\mathbf{3}^{(1)}\stackrel{u_{2}}{\longleftrightarrow} \mathbf{3}^{(3)},~~(\mathbf{3}^{(1)})^{*}\stackrel{u_{2}}{\longleftrightarrow} (\mathbf{3}^{(3)})^{*}.
\end{equation}
They could be realised through the mappings~\cite{42}:
\begin{equation}
~a\stackrel{u_{2}}{\longrightarrow} ca^{2}c,~~b\stackrel{u_{2}}{\longrightarrow} c^{2}a^{2}cb^{3},~~c\stackrel{u_{2}}{\longrightarrow} ca^{2}.
\end{equation}
The actions of the outer automorphism $u_{1}u_{2}$ read
\begin{equation}
1C^{3}_{2}\stackrel{u_{1}u_{2}}{\longleftrightarrow} 1C^{3}_{3},~~9C_{7}^{6}\stackrel{u_{1}u_{2}}{\longleftrightarrow} 9C_{8}^{6},~~9C_{9}^{4}\stackrel{u_{1}u_{2}}{\longleftrightarrow} 9C_{12}^{4},~~9C_{10}^{12}\stackrel{u_{1}u_{2}}{\longleftrightarrow} 9C_{14}^{12},~~9C_{11}^{12}\stackrel{u_{1}u_{2}}{\longleftrightarrow} 9C_{13}^{12},
\end{equation}
\begin{equation}
\mathbf{1}^{(1)}\stackrel{u_{1}u_{2}}{\longleftrightarrow} \mathbf{1}^{(3)}=(\mathbf{1}^{(1)})^{*},
~~\mathbf{3}^{(0)}\stackrel{u_{1}u_{2}}{\longleftrightarrow} (\mathbf{3}^{(0)})^{*},~~\mathbf{3}^{(2)}\stackrel{u_{1}u_{2}}{\longleftrightarrow} (\mathbf{3}^{(2)})^{*},~~\mathbf{3}^{(1)}\stackrel{u_{1}u_{2}}{\longleftrightarrow} (\mathbf{3}^{(1)})^{*},~~\mathbf{3}^{(3)}\stackrel{u_{1}u_{2}}{\longleftrightarrow} (\mathbf{3}^{(3)})^{*}.
\end{equation}
They could be realised through the mappings:
\begin{equation}
\label{eq:24}
~a\stackrel{u_{1}u_{2}}{\longrightarrow} a,~~b\stackrel{u_{1}u_{2}}{\longrightarrow} b^{-1},~~c\stackrel{u_{1}u_{2}}{\longrightarrow} c^{-1}.
\end{equation}
As we can see, $u_{1}u_{2}$ interchanges all representations with their complex conjugations~\cite{42}.

\subsection{Semidirect approach of GCP }
\subsubsection{General GCP compatible with $\Sigma(36\times3)$}
We consider a theory of leptons which satisfies the symmetry $G_{f}\rtimes H_{CP}$, where $G_{f}$ is the flavor group $\Sigma(36\times3)$, $H_{CP}$
is the group of GCP. The element of $H_{CP}$  satisfies the consistent condition~\cite{23}
\begin{equation}
\label{eq:25}
X(w)\rho^{*}(g)X^{-1}(w)=\rho(w(g)), ~~with ~w \in H_{CP},~g, w(g) \in G_{f},
\end{equation}
where $X$, $\rho$ denote the representation of $H_{CP}$ and that of $G_{f}$ respectively. This condition reveals that the CP transformation is a class-inverting automorphism (CIA) of the flavor group~\cite{23,45}. For the outer automorphism group $K_{4}=\{~id, u_{1}, u_{2}, u_{1}u_{2}~\}$, the unique CIA is $u_{1}u_{2}$. According to the action of $u_{1}u_{2}$ (see Eq.(\ref{eq:24})) and the 3-dimensional representation of the generators of $\Sigma(36\times3)$ (see Eq.(\ref{eq:3})), $X(u_{1}u_{2})$ satisfies the equations as follows:
\begin{equation}
\begin{array}{c}
X(u_{1}u_{2})\rho^{*}(a) X^{-1}(u_{1}u_{2})=\rho(a)=\rho^{*}(a),\\
X(u_{1}u_{2})\rho^{*}(b) X^{-1}(u_{1}u_{2})=\rho^{-1}(b)=\rho^{*}(b),\\
X(u_{1}u_{2})\rho^{*}(c) X^{-1}(u_{1}u_{2})=\rho^{-1}(c)=\rho^{*}(c).
\end{array}
\end{equation}
The solution (up to a global phase) is
\begin{equation}
X(u_{1}u_{2})= \left(
                 \begin{array}{ccc}
                   1 & 0 & 0 \\
                   0 & 1 & 0 \\
                   0 & 0 & 1 \\
                 \end{array}
               \right).
\end{equation}
Therefore, the general GCP which is compatible with the group $\Sigma(36\times3)$ in the 3-dimensional representation is of the form
\begin{equation}
X(w)=\rho_{3}(g)X(u_{1}u_{2})=\rho_{3}(g)=\rho(c^{i}a^{j}b^{k})~~with~~ i, j=0, 1, 2,~ k=0, 1 ,2, 3.
\end{equation}
Note that the global phase $\omega^{i}$ is not considered here which corresponds to the element of the centre i.e. $Z_{3}$.
Furthermore, the CP transformation $X(w)$ should be a symmetric and unitary matrix~\cite{24}, i.e.
\begin{equation}
X(w)=X^{T}(w),~~XX^{*}=E.
 \end{equation}
This type of CP transformation is the so-called Bickerstaff-Damhus automorphism~\cite{44,45}. And $X(w)$ could be decomposed as $X(w)=\Omega_{w}\Omega_{w}^{T}$.
On the base of this decomposition, $X(w)$ could be transformed to a unit matrix in the new basis called CP basis, namely
\begin{equation}
\Omega_{w}^{+}X(w)\Omega_{w}^{*}=diag(1, 1, 1).
\end{equation}

\subsubsection{Residual flavor symmetries with GCP }
After leptons obtain masses through the vacuum expectation values of scalars, the original symmetry $G_{f}\rtimes H_{CP}$ is broken to $G_{e}\rtimes H^{e}_{CP}$ in the charged lepton sector and $G_{\nu}\rtimes H^{\nu}_{CP}$ in the neutrino sector. In the semidirect method, $G_{\nu}$ is $Z_{2}$. So $G_{\nu}\rtimes H^{\nu}_{CP}$ is reduced to $Z_{2}\times H^{\nu}_{CP}$. For the Majorana neutrinos mass matrix, we have following relations:
\begin{equation}
\rho^{T}(g_{\nu})m_{\nu}\rho(g_{\nu})=m_{\nu},~~X_{\nu}^{T}(w)m_{\nu}X_{\nu}(w)=m^{*}_{\nu},~~with~~ X_{\nu}(w)\rho^{*}(g_{\nu})X_{\nu}^{-1}(w)=\rho(g_{\nu}),
\end{equation}
where $g_{\nu}\in Z_{2}$, $w\in H^{\nu}_{CP}$.
The unitary matrix $U_{\nu}$ fulfilling $U_{\nu}^{T}m_{\nu}U_{\nu}= diag (m_{1},~m_{2},~m_{3})$, could be expressed
as~\cite{24}
\begin{equation}
U_{\nu}=\Omega_{\nu} R_{\nu ij}(\theta)P_{\nu},
\end{equation}
where $R_{\nu ij}(\theta)$ is a rotation matrix in the plane with $ij=12, 23, 13$. $P_{\nu}$ is a phase matrix which coordinates the sign of $m_{i}$, i.e.
\begin{equation}
\label{eq:36}
P_{\nu}=diag(1,~ i^{k_{1}}, ~i^{k_{2}}), ~~with ~~k_{1}, k_{2}=0, 1, 2, 3.
\end{equation}

In the charged lepton sector, we consider two nontrivial cases, namely\\
Case A:
\begin{equation}
G_{e}\rtimes H^{e}_{CP}= Z_{n}\rtimes H^{e}_{CP}, ~~with ~n\geq 3,
\end{equation}
Case B:
\begin{equation}
G_{e}\rtimes H^{e}_{CP}= Z_{2}\times H^{e}_{CP}.
\end{equation}
In Case A, the mass matrix of the charged leptons is constrained by the residual symmetries as
\begin{equation}
\label{eq:39}
\rho^{+}(g_{e})m_{e}m^{+}_{e}\rho(g_{e})=m_{e}m^{+}_{e},~~ X_{e}^{+}(w)(m_{e}m^{+}_{e})X_{e}(w)=(m_{e}m^{+}_{e})^{*}, ~~with ~ g_{e} \in Z_{n},~ w\in H^{e}_{CP} .
\end{equation}
And the unitary matrix $U_{e}$ which fulfills $U_{e}^{+}m_{e}m^{+}_{e}U_{e}= diag (m^{2}_{e},~m^{2}_{\mu},~m^{2}_{\tau})$, could be obtained from the diagonalization of $\rho(g_{e})$, namely
\begin{equation}
U^{+}_{e}\rho(g_{e})U_{e}=\rho ^{d}(g_{e}),
\end{equation}
where $\rho ^{d}(g_{e})$ is diagonal. $U_{e}$ is fixed by the residual flavor group up to permutations of columns and nonphysical phases. So the effect of the residual CP symmetry is not considered in this case in the following sections. \\
In Case B, the constraints are same as those in Eq.~(\ref{eq:39}) but
with $g_{e}\in Z_{2}$.
The matrix $U_{e}$ could be expressed as
 \begin{equation}
\label{eq:43}
U_{e}=\Omega_{e} R_{e ij}(\theta_{2}),
\end{equation}
where $R_{e ij}$ is also a rotation matrix. Different from the case of $U_{\nu}$, the nonphysical phase matrix is not included in $U_{e}$.\\

\subsubsection{Similarity transformation}
If a combination of residual flavor symmetries with the GCP, i.e., $(Z_{n(e)}, ~Z_{2(\nu)}, ~X_{\nu})$ in Case A or $(Z_{2(e)},~X_{e}, ~Z_{2(\nu)}, ~X_{\nu})$ in Case B is given, the leptonic mixing matrix $U_{PMNS}=U^{+}_{e}U_{\nu}$ could be obtained (up to permutations of rows or columns).
If two combinations are related by a similarity transformation, namely~\cite{24}
\begin{equation}
\label{eq:44}
\rho^{\prime}(g_{\alpha})=\Omega_{0}\rho(g_{\alpha})\Omega^{+}_{0},~~
X^{\prime}_{\beta}=\Omega_{0}X_{\beta}\Omega^{T}_{0},
\end{equation}
with $\alpha=e, \nu$, $\beta=\nu$ in Case A, ~$\beta= e,~\nu$ in Case B,
they would correspond to the same mixing matrix of leptons. Although there are 13 $Z_{3}$ and 9 $Z_{4}$ subgroups of $\Sigma(36\times3)$, we could just consider the subgroup  generated by a representative in the conjugacy class.
So in Case A, we consider the combination of $Z_{n(e)}$ listed in Table~\ref{tab:1} and $(Z_{2(\nu)}, ~X_{\nu})$ in Table~\ref{tab:2}. Furthermore, the $Z_{6}$ and $Z_{12}$ subgroups in the 3-dimensional representation are different from the $Z_{2}$ and $Z_{3}$ subgroups just because of the factor $\omega$. They don't bring new mixing patterns.  According to the same reason, in Case B we could fix $Z_{2(e)}$ on $Z^{b^{2}}_{2}$ and consider the combination $(Z^{b^{2}}_{2(e)},~X_{e}, ~Z_{2(\nu)}, ~X_{\nu})$.

\section{Results}

\subsection{Combinations in Case A}
In Case A, the  diagonalization matrix $U_{e}$ is completely determined by $Z_{n (e)}$. And a column of the matrix $U_{\nu}$ is determined by $(Z_{2 (\nu)},~X_{\nu})$. Finally, a column of the leptonic mixing matrix $U_{PMNS}$ is fixed. There is a free parameter to coordinate the mixing patterns. Various of nonequivalent combinations of $(Z_{n(e)}, ~Z_{2(\nu)}, ~X_{\nu})$ are surveyed. We find that neither of them could accommodate the fit data of neutrinos~\cite{46} at the $3\sigma$ level. This observation could be seen from the fixed column of the mixing matrix. For the combinations $(Z^{c}_{3(e)}, ~Z_{2(\nu)}, ~X_{\nu})$ and
$(Z^{ca}_{3(e)}, ~Z_{2(\nu)}, ~X_{\nu})$, the magnitude of the fixed column vector (up to permutations of rows) is
\begin{equation}
\label{eq:45}
|U_{PMNS}(\alpha, i)|=\left(
                        \begin{array}{ccc}
                          0, & \frac{\sqrt{2}}{2}, & \frac{\sqrt{2}}{2} \\
                        \end{array}
                      \right)^{T}, ~with ~\alpha=e,~\mu,~\tau.
\end{equation}
For the combinations $(Z^{b}_{4(e)}, ~Z_{2(\nu)}, ~X_{\nu})$, the magnitude of the column vector includes two cases.
If $Z_{2(\nu)}=Z^{b^{2}}_{2(\nu)}$, we have
\begin{equation}
|U_{PMNS}(\alpha, i)|=\left(
                        \begin{array}{ccc}
                         0, &  0, & 1 \\
                        \end{array}
                      \right)^{T}.
\end{equation}
Otherwise,
\begin{equation}
|U_{PMNS}(\alpha, i)|=\left(
                        \begin{array}{ccc}
                         \frac{ 1}{2}\sqrt{\frac{2\sqrt{3}-3}{\sqrt{3}-1}}, &  \frac{ 1}{2}\sqrt{\frac{2\sqrt{3}+3}{\sqrt{3}+1}}, & \frac{1}{2} \\
                        \end{array}
                      \right)^{T}\simeq\left(
                        \begin{array}{ccc}
                        0.398, &  0.769, & 0.5 \\
                        \end{array}
                      \right)^{T}.
\end{equation}
 Anyway, neither of these columns is in the $3\sigma$ range of the fit date of neutrinos~\cite{46}
\begin{equation}
|U_{PMNS}|=
\left(
\begin{array}{ccc}
 0.800\rightarrow0.844 & 0.515\rightarrow0.581 & 0.139\rightarrow0.155 \\
 0.229\rightarrow0.516 & 0.438\rightarrow0.699 & 0.614\rightarrow0.790 \\
0.249\rightarrow0.528 &0.462\rightarrow0.715 & 0.595\rightarrow0.776
\end{array}
\right).
\end{equation}
Similar observations were obtained in Ref.~\cite{42}.

\begin{table}
\caption{~$Z_{n}$ $(n\geq3)$ subgroup of $\Sigma(36\times3)$ generated by a representative in the conjugacy class. $Z^{g_{\alpha}}_{n}$ denotes the group $Z_{n}$ generated by the element $g_{\alpha}$. Other subgroups which could be obtained by the group conjugation or the factor $\omega=e^{i2\pi/3}$ are not listed here.}
\label{tab:1}
\begin{tabular}{||c|c|c|c||}
\noalign{\smallskip}\hline
~~~~Z$_{n}$~~~~&~~~~Z$^{c}_{3}$~~~~ &~~~~Z$^{ca}_{3}$~~~~& ~~~~Z$^{b}_{4}$~~~~ \\[0.5ex]\hline
\noalign{\smallskip}\noalign{\smallskip}
\noalign{\smallskip}
\end{tabular}
\vspace*{0.5cm}
\end{table}

\begin{table}
\caption{~Generators of $Z_{2}$ subgroups of $\Sigma(36\times3)$ with the corresponding GCP in the 3-dimensional representation.}
\label{tab:2}       % Give a unique label
% For LaTeX tables use
\begin{tabular}{||c|c|c|c|c||}
\noalign{\smallskip}\hline
$\rho(g_{\alpha})$ &~~~~X$^{(1)}~~~~$&~~~~ X$^{(2)}~~~~$&X$^{(3)}$ &X$^{(4)}$\\[0.5ex]\hline
\noalign{\smallskip}\noalign{\smallskip}\hline
$\rho(b^{2}):\left(
                         \begin{array}{ccc}
                           -1 & 0 & 0 \\
                           0 & 0 & -1 \\
                           0 & -1& 0 \\
                         \end{array}
                       \right)$ & $\left(
                                          \begin{array}{ccc}
                                            1 & 0 & 0 \\
                                            0 & 1 & 0 \\
                                            0 & 0 & 1 \\
                                          \end{array}
                                        \right)$ &$\left(
                                          \begin{array}{ccc}
                                            -1 & 0 & 0 \\
                                            0 & 0 & -1 \\
                                            0 & -1 & 0 \\
                                          \end{array}
                                        \right)$ & $\frac{1}{\sqrt{3}i}
\left(
  \begin{array}{ccc}
   1 & 1 & 1 \\
    1 & \omega & \omega^{2} \\
    1 & \omega^{2} & \omega \\
  \end{array}
\right)$& $\frac{-1}{\sqrt{3}i}
\left(
  \begin{array}{ccc}
   1 & 1 & 1 \\
    1 & \omega^{2} & \omega \\
    1 & \omega & \omega^{2} \\
  \end{array}
\right)$  \\\hline
$\rho(ab^{2}):\left(
                                          \begin{array}{ccc}
                                            0 & 0 & -1 \\
                                            0 & -1 & 0 \\
                                            -1 & 0 & 0 \\
                                          \end{array}
                                        \right)$&$\left(
                                          \begin{array}{ccc}
                                            1 & 0 & 0 \\
                                            0 & 1 & 0 \\
                                            0 & 0 & 1 \\
                                          \end{array}
                                        \right)$ &  $\left(
                                          \begin{array}{ccc}
                                            0 & 0 & -1 \\
                                            0 & -1 & 0 \\
                                            -1 & 0 & 0 \\
                                          \end{array}
                                        \right)$ & $\frac{1}{\sqrt{3}}
\left(
  \begin{array}{ccc}
   e^{i\pi/6} & -i & e^{i5\pi/6} \\
    -i & -i& -i \\
    e^{i5\pi/6} & -i & e^{i\pi/6} \\
  \end{array}
\right)$&$\frac{1}{\sqrt{3}}
\left(
  \begin{array}{ccc}
   e^{-i\pi/6} & i & e^{-i5\pi/6} \\
    i & i& i \\
    e^{-i5\pi/6} & i & e^{-i\pi/6} \\
  \end{array}
\right)$ \\\hline
$\rho(a^{2}b^{2}):\left(
                                          \begin{array}{ccc}
                                            0 & -1 & 0 \\
                                            -1 & 0 & 0 \\
                                            0 & 0 & -1 \\
                                          \end{array}
                                        \right)$ &$\left(
                                          \begin{array}{ccc}
                                            1 & 0 & 0 \\
                                            0 & 1 & 0 \\
                                            0 & 0 & 1 \\
                                          \end{array}
                                        \right)$&$\left(
                                          \begin{array}{ccc}
                                            0 & -1 & 0 \\
                                            -1 & 0 & 0 \\
                                            0 & 0 & -1 \\
                                          \end{array}
                                        \right)$&$\frac{1}{\sqrt{3}}
\left(
  \begin{array}{ccc}
   e^{i\pi/6} & e^{i5\pi/6} & -i \\
    e^{i5\pi/6} &e^{i\pi/6} & -i \\
    -i & -i & -i \\
  \end{array}
\right)$& $\frac{1}{\sqrt{3}}
\left(
  \begin{array}{ccc}
   e^{-i\pi/6} & e^{-i5\pi/6} & i \\
    e^{-i5\pi/6} &e^{-i\pi/6} & i \\
    i & i & i \\
  \end{array}
\right)$ \\\hline
$\rho(cb^{2}):\left(
                         \begin{array}{ccc}
                           -1 & 0 & 0 \\
                           0 & 0 & -\omega \\
                           0 & -\omega^{2}& 0 \\
                         \end{array}
                       \right)$& $
\left(
  \begin{array}{ccc}
   1 & 0 & 0\\
    0 &\omega & 0 \\
    0 & 0 & \omega^{2} \\
  \end{array}
\right)$ &  $\left(
                                          \begin{array}{ccc}
                                            -1 & 0 & 0 \\
                                            0 & 0 & -1 \\
                                            0 & -1 & 0 \\
                                          \end{array}
                                        \right)$ & $\frac{1}{\sqrt{3}}
\left(
  \begin{array}{ccc}
   -i & e^{i5\pi/6} & e^{i\pi/6} \\
    e^{i5\pi/6} &e^{i5\pi/6} &e^{i5\pi/6} \\
    e^{i\pi/6} & e^{i5\pi/6} & -i \\
  \end{array}
\right)$&$\frac{1}{\sqrt{3}}
\left(
  \begin{array}{ccc}
   i & e^{-i\pi/6} & e^{-i5\pi/6} \\
    e^{-i\pi/6} &i &e^{-i5\pi/6} \\
    e^{-i5\pi/6} & e^{-i5\pi/6} & e^{-i5\pi/6} \\
  \end{array}
\right)$~ \\\hline
~$\rho(c^{2}b^{2}):\left(
                         \begin{array}{ccc}
                           -1 & 0 & 0 \\
                           0 & 0 & -\omega^{2} \\
                           0 & -\omega& 0 \\
                         \end{array}
                       \right)$& $
\left(
  \begin{array}{ccc}
   1 & 0 & 0\\
    0 &\omega^{2} & 0 \\
    0 & 0 & \omega\\
  \end{array}
\right)$ &  $\left(
\begin{array}{ccc}
                                            -1 & 0 & 0 \\
                                            0 & 0 & -1 \\
                                            0 & -1 & 0 \\
                                          \end{array}
                                        \right)$&  $\frac{1}{\sqrt{3}}
\left(
  \begin{array}{ccc}
   -i & e^{i\pi/6} & e^{i5\pi/6} \\
    e^{i\pi/6} &-i &e^{i5\pi/6} \\
    e^{i5\pi/6} & e^{i5\pi/6} & e^{i5\pi/6} \\
  \end{array}
\right)$&$\frac{1}{\sqrt{3}}
\left(
  \begin{array}{ccc}
   i & e^{-i5\pi/6} & e^{-i\pi/6} \\
    e^{-i5\pi/6} &e^{-i5\pi/6} &e^{-i5\pi/6} \\
    e^{-i\pi/6} & e^{-i5\pi/6} & i \\
  \end{array}
\right)$ \\\hline
$\omega\rho(ca^{2}b^{2}):\left(
                         \begin{array}{ccc}
                           0 & -\omega & 0 \\
                           -\omega^{2} & 0 & 0 \\
                           0 & 0& -1 \\
                         \end{array}\right)$~&$
\left(
  \begin{array}{ccc}
   1 & 0 & 0\\
    0 &\omega & 0 \\
    0 & 0 & \omega^{2} \\
  \end{array}
\right)$ &  $e^{i\pi/3}\left(
                                          \begin{array}{ccc}
                                            0 & 1 & 0 \\
                                            1 & 0 & 0 \\
                                            0 & 0 & 1 \\
                                          \end{array}
                                        \right)$ &  $\frac{1}{\sqrt{3}}
\left(
  \begin{array}{ccc}
   e^{i\pi/6} & e^{i\pi/6} & e^{i\pi/6} \\
    e^{i\pi/6} &e^{i5\pi/6} &-i \\
    e^{i\pi/6} & -i & e^{i5\pi/6} \\
  \end{array}
\right)$&$\frac{1}{\sqrt{3}}
\left(
  \begin{array}{ccc}
   e^{-i\pi/6} & i & e^{-i5\pi/6} \\
    i & i& i \\
    e^{-i5\pi/6} & i & e^{-i\pi/6} \\
  \end{array}
\right)$\\\hline
~$\omega^{2}\rho(c^{2}a^{2}b^{2}):\left(
                         \begin{array}{ccc}
                           0 & -\omega^{2} & 0 \\
                           -\omega & 0 & 0 \\
                           0 & 0& -1 \\
                         \end{array}\right)$& $
\left(
  \begin{array}{ccc}
  \omega & 0 & 0\\
    0 &1 & 0 \\
    0 & 0 & \omega^{2} \\
  \end{array}
\right)$ & $e^{i\pi/3}\left(
                                          \begin{array}{ccc}
                                            0 & 1 & 0 \\
                                            1 & 0 & 0 \\
                                            0 & 0 & 1 \\
                                          \end{array}
                                        \right)$ &  $\frac{1}{\sqrt{3}}
\left(
  \begin{array}{ccc}
   e^{i5\pi/6} &  e^{i\pi/6}& -i \\
    e^{i\pi/6} & e^{i\pi/6}& e^{i\pi/6} \\
    -i & e^{i\pi/6} & e^{i5\pi/6} \\
  \end{array}
\right)$&$\frac{1}{\sqrt{3}}
\left(
  \begin{array}{ccc}
   i &  i& i \\
   i & e^{-i\pi/6}& e^{-i5\pi/6} \\
    i & e^{-i5\pi/6} & e^{-i\pi/6} \\
  \end{array}
  \right)$~~ \\\hline
~~~~$\omega\rho(c^{2}ab^{2}):\left(
                         \begin{array}{ccc}
                           0 & 0 & -\omega \\
                           0 & -1 & 0 \\
                           -\omega^{2} & 0& 0 \\
                         \end{array}\right)$& $
\left(
  \begin{array}{ccc}
   1 & 0 & 0\\
    0 &\omega^{2} & 0 \\
    0 & 0 & \omega \\
  \end{array}
\right)$ &  $
e^{i\pi/3}\left(
  \begin{array}{ccc}
   0 & 0 & 1\\
    0 &1 & 0 \\
    1 & 0 & 0 \\
  \end{array}
\right)$ &$\frac{1}{\sqrt{3}}
\left(
  \begin{array}{ccc}
   e^{i\pi/6} &  e^{i\pi/6}& e^{i\pi/6} \\
   e^{i\pi/6} & e^{i5\pi/6}& -i \\
    e^{i\pi/6} & -i & e^{i5\pi/6} \\
  \end{array}
  \right)$&$\frac{1}{\sqrt{3}}
\left(
  \begin{array}{ccc}
   e^{-i\pi/6} &  e^{-i5\pi/6}& i \\
   e^{-i5\pi/6} & e^{-i\pi/6}& i \\
    i & i & i \\
  \end{array}
  \right)$ \\\hline
$\omega^{2}\rho(cab^{2}):\left(
                         \begin{array}{ccc}
                           0 & 0 & -\omega^{2} \\
                           0 & -1 & 0 \\
                           -\omega & 0& 0 \\
                      \end{array}\right)$& $
\left(
  \begin{array}{ccc}
    \omega & 0 & 0\\
    0 &\omega^{2} & 0 \\
    0 & 0 &1\\
  \end{array}
\right)$ & $
e^{i\pi/3}\left(
  \begin{array}{ccc}
    0 & 0 & 1\\
    0 &1 & 0 \\
    1 & 0 &0\\
  \end{array}
\right)$&  $\frac{1}{\sqrt{3}}
\left(
  \begin{array}{ccc}
   e^{i5\pi/6} &  -i& e^{i\pi/6} \\
   -i & e^{i5\pi/6}& e^{i\pi/6}\\
    e^{i\pi/6} & e^{i\pi/6} & e^{i\pi/6} \\
  \end{array}
  \right)$&$\frac{1}{\sqrt{3}}
\left(
  \begin{array}{ccc}
   i &  i& i \\
   i & e^{-i\pi/6}& e^{-i5\pi/6}\\
   i & e^{-i5\pi/6} & e^{-i\pi/6} \\
  \end{array}
  \right)$~~ \\\hline
\noalign{\smallskip}
\end{tabular}
% Or use
\vspace*{0.5cm}  % with the correct table height
\end{table}

\subsection{Combinations in Case B}
\subsubsection{Viable combinations }
In Case B, on the basis of the similarity transformation, the residual symmetry in the charged lepton sector is fixed as ($Z^{b^{2}}_{2}$, $X_{e}=E,~ \rho(b)$). Note that $X^{(1)}$ and $X^{(2)}$ for a given group $Z_{2}$ in Table~\ref{tab:2} correspond to the same mixing pattern. So do $X^{(3)}$ and $X^{(4)}$. The verifications of equivalence of $X^{(i)}$ are given in the appendix. Thus, we just consider two cases of  $X_{e}$ for a fixed $Z_{2}$. In the sector of neutrinos, the residual symmetry could be expressed as ($Z^{g_{\alpha}}_{2}$, $X_{\nu}=X_{\nu}^{(1)},X_{\nu}^{(3)}$). The 3-dimensional representations of generators of $Z_{2}$ subgroups and those of GCP are listed in Table~\ref{tab:2}. In total, there are 36 different combinations of residual symmetries in Case B. We perform a $\chi^{2}$ analysis on these combinations. The $\chi^{2}$ function is defined as
\begin{equation}
\chi^{2}=\sum_{ij=13,23,12}(\frac{\sin^{2}\theta_{ij}-(\sin^{2}\theta_{ij})^{bf}}{\sigma_{ij}})^{2},
\end{equation}
where $(\sin^{2}\theta_{ij})^{bf}$ are best fit values from Ref.~\cite{46}, $\sigma_{ij}$ is the 1$\sigma$ error.
Only 2 types of combinations can accommodate the global fit data at 3$\sigma$ level~\cite{46}, namely\\
Type I:
\begin{equation}
\label{eq:49}
\begin{array}{c}
 (Z^{ b^{2}}_{2(e)},E, ~Z^{a^{2}b^{2}}_{2(\nu)},~X^{(1)}= E),(Z^{ b^{2}}_{2(e)},~E, ~Z^{ab^{2}}_{2(\nu)},~X^{(1)}=E),\\
 (Z^{ b^{2}}_{2(e)},~E, ~Z^{cb^{2}}_{2(\nu)},~ X^{(1)}=\rho(c)),~(Z^{ b^{2}}_{2(e)},E, ~Z^{c^{2}b^{2}}_{2(\nu)},~ X^{(1)}=\rho^{*}(c)),
\end{array}
\end{equation}
Type V:
\begin{equation}
\label{eq:50}
(Z^{ b^{2}}_{2(e)},~E, ~ Z^{a^{2}b^{2}}_{2(\nu)}, ~X^{(3)}= \rho(ab^{3}ab^{2})),~~(Z^{ b^{2}}_{2(e)},~E, ~Z^{cb^{2}}_{2(\nu)},~ X^{(3)}=\rho(c^{2}bc^{2})),~~(Z^{ b^{2}}_{2(e)},~E, ~Z^{c^{2}b^{2}}_{2(\nu)},~ X^{(3)}=\rho(cbc)).
\end{equation}
The mixing patterns of the combinations of Type I are equivalent. So are those of
combinations of Type V. The transformation relation for the equivalence is given in the appendix. Thus, we just show the mixing pattern of the combination $(Z^{ b^{2}}_{2(e)},~E, ~Z^{a^{2}b^{2}}_{2(\nu)},~ E)$ of Type I, and that of the combination $(Z^{ b^{2}}_{2(e)},~E, ~ Z^{a^{2}b^{2}}_{2(\nu)}, ~ \rho(ab^{3}ab^{2}))$ of Type V.

\subsubsection{Analytical expressions for the viable mixing patterns}
In the sector of the charged leptons, the residual symmetry is $Z^{ b^{2}}_{2}\times X_{e},$ with $X_{e}=E$.
The unit matrix could be decomposed as $E=\Omega_{e}\Omega_{e}^{T}$, with $\Omega_{e}^{+}\rho(b^{2})\Omega_{e}=diag(-1,~-1,~1)$. We could choose $\Omega_{e}$ as
\begin{equation}
\Omega_{e I}=\left(
             \begin{array}{ccc}
              0 & 1 & 0 \\
                \frac{\sqrt{2}}{2} & 0 &  -\frac{\sqrt{2}}{2} \\
              \frac{\sqrt{2}}{2} & 0 & \frac{\sqrt{2}}{2} \\
             \end{array}
           \right).
\end{equation}
The unitary matrix $U_{e}$ could be expressed as
\begin{equation}
U_{e}=\Omega_{eI} R_{e 12}(\theta_{2})=
\left(
        \begin{array}{ccc}
          -\sin\theta_{2} & \cos\theta_{2} & 0 \\
          \frac{\sqrt{2}}{2}\cos\theta_{2} & \frac{\sqrt{2}}{2}\sin\theta_{2} & - \frac{\sqrt{2}}{2} \\
          \frac{\sqrt{2}}{2}\cos\theta_{2} & \frac{\sqrt{2}}{2}\sin\theta_{2} & \frac{\sqrt{2}}{2} \\
        \end{array}
      \right),
\end{equation}
where the rotation matrix $R_{12}(\theta_{2})$ reads
\begin{equation}
R_{12}(\theta_{2})=\left(
        \begin{array}{ccc}
         \cos\theta_{2} & \sin\theta_{2} & 0 \\
          -\sin\theta_{2}  & \cos\theta_{2} & 0 \\
          0 & 0 & 1 \\
        \end{array}
      \right).
\end{equation}

In the neutrino sector, the residual flavor symmetry with the GCP is $Z^{ a^{2}b^{2}}_{2}\times X_{\nu},$ with $X_{\nu}=E, \rho(ab^{3}ab^{2})$. For $X_{\nu}=E$, we can choose
\begin{equation}
\Omega_{\nu I}=\left(
             \begin{array}{ccc}
               - \frac{\sqrt{2}}{2} & 0 & - \frac{\sqrt{2}}{2} \\
               - \frac{\sqrt{2}}{2} & 0 &  \frac{\sqrt{2}}{2} \\
              0 & -1 & 0 \\
             \end{array}
           \right).
\end{equation}
And for $X_{\nu}=\rho(ab^{3}ab^{2})$, we choose
\begin{equation}
\label{eq:55}
\Omega_{\nu V}=\left(
             \begin{array}{ccc}
               - \frac{e^{i\pi/4}\cos\theta_{1}}{\sqrt{2}} &  - \frac{e^{3i\pi/4}\sin\theta_{1}}{\sqrt{2}} & - \frac{\sqrt{2}}{2} \\
                - \frac{e^{i\pi/4}\cos\theta_{1}}{\sqrt{2}} & - \frac{e^{3i\pi/4}\sin\theta_{1}}{\sqrt{2}} &  \frac{\sqrt{2}}{2} \\
                e^{i\pi/4}\sin\theta_{1}& - e^{3i\pi/4}\cos\theta_{1}& 0 \\
             \end{array}
           \right), ~~with~ \theta_{1}=\frac{1}{2}\arctan\sqrt{2}.
\end{equation}
And the mixing matrix $U_{\nu}$ is obtained according to the equation $U_{\nu}=\Omega_{\nu \gamma} R_{\nu 12}(\theta)P_{\nu}$, with $\gamma=I, V$.
Then the lepton mixing matrix $U_{PMNS}$ is determined up to permutations of rows or columns with the matrices
\begin{equation}
S_{12}=\left(
 \begin{array}{ccc}
0 & 1 &0 \\
1 &0 & 0 \\
0 & 0 &1 \\
\end{array}
\right),
S_{23}=
\left(
\begin{array}{ccc}
 1 & 0 &0 \\
0 & 0 & 1 \\
0 & 1 &0 \\
\end{array}
\right),
S_{13}=\left(
\begin{array}{ccc}
 0 & 0 &1 \\
  0 &1 & 0 \\
   1 & 0 &0 \\
 \end{array}
 \right).
\end{equation}
In detail, for the combination of Type I, we obtain the viable mixing pattern I1:
\begin{equation}
\begin{aligned}
~~U_{I1}&=U^{+}_{e}(\theta_{2})\Omega_{\nu I} R_{12}(\theta)P_{\nu}S_{23}\\
&=
\left(
  \begin{array}{ccc}
    -\frac{1}{2}\cos\theta\cos\theta_{2}+\frac{\sqrt{2}}{2}\sin(\theta+\theta_{2}) & \frac{ \sqrt{3}}{2}\cos(\theta_{2}-2\theta_{1}) &  -\frac{1}{2}\sin\theta\cos\theta_{2}-\frac{\sqrt{2}}{2}\cos(\theta+\theta_{2})\\
     -\frac{1}{2}\cos\theta\sin\theta_{2}-\frac{\sqrt{2}}{2}\cos(\theta+\theta_{2}) &  \frac{ \sqrt{3}}{2}\sin(\theta_{2}-2\theta_{1}) & -\frac{1}{2}\sin\theta\sin\theta_{2}-\frac{\sqrt{2}}{2}\sin(\theta+\theta_{2}) \\
   \frac{ \sqrt{3}}{2}\cos(\theta-2\theta_{1}) & -\frac{1 }{2}&  \frac{ \sqrt{3}}{2}\sin(\theta-2\theta_{1})\\
  \end{array}
\right)P^{\prime}_{\nu},
\end{aligned}
\end{equation}
where $P^{\prime}_{\nu}=S_{23}P_{\nu}S_{23}$,
and its nonequivalent permutations
\begin{equation}
\begin{aligned}
I2:~~~~~U_{I2}&=S_{23}U_{I1},\\
I3:~~~~~U_{I3}&=U_{I1}S_{23}S_{13},\\
I4:~~~~~U_{I4}&=S_{13}S_{23}U_{I1}S_{23}S_{13}.
\end{aligned}
\end{equation}
Compared with the parametrization of the lepton mixing matrix as follow
\begin{equation}
U_{PMNS}=
\left(
\begin{array}{ccc}
 c_{12}c_{13} & s_{12}c_{13} & s_{13}e^{-i\delta} \\
 -s_{12}c_{23}-c_{12}s_{13}s_{23}e^{i\delta} & c_{12}c_{23}-s_{12}s_{13}s_{23}e^{i\delta} & c_{13}s_{23} \\
s_{12}s_{23}-c_{12}s_{13}c_{23}e^{i\delta} & -c_{12}s_{23}-s_{12}s_{13}c_{23}e^{i\delta} & c_{13}c_{23}
\end{array}
\right)
\left(
\begin{array}{ccc}
1 & 0 & 0 \\
 0 & e^{i\alpha/2} & 0 \\
0 & 0 & e^{i(\beta/2+\delta)}
\end{array}
\right),
\end{equation}
where $s_{ij}\equiv\sin{\theta_{ij}}$, $c_{ij}\equiv\cos{\theta_{ij}}$, $\delta$ is the Dirac CP-violating phase, $\alpha$ and $\beta$ are Majorana Phases, the mixing angles are obtained as follows\\
I1:
\begin{equation}
 \begin{aligned}
 \sin^{2}\theta_{13}&=\frac{1}{2}[\cos(\theta+\theta_{2})+\frac{1}{\sqrt{2}}\sin\theta\cos\theta {_2}]^2,\\
 \sin^{2}\theta_{23}&=\frac{1}{2}[\frac{1} {\sqrt{2}}\sin\theta\sin\theta {_2}+\sin(\theta+\theta_{2})]^2/(1-\sin^{2}\theta_{13}),\\
 \sin^{2}\theta_{12}&=\frac{3}{4}\cos^{2}(\theta_{2}-2\theta_{1})/(1-\sin^{2}\theta_{13});
\end{aligned}
\end{equation}
I2:
\begin{equation}
\label{eq:29}
 \begin{aligned}
 \sin^{2}\theta_{13}&=\frac{1}{2}[\cos(\theta+\theta_{2})+\frac{1}{\sqrt{2}}\sin\theta\cos\theta {_2}]^2,\\
 \sin^{2}\theta_{23}&=\frac{3}{4}\sin^{2}(\theta-2\theta_{1})/(1-\sin^{2}\theta_{13}),\\
 \sin^{2}\theta_{12}&=\frac{3}{4}\cos^{2}(\theta_{2}-2\theta_{1})/(1-\sin^{2}\theta_{13});
\end{aligned}
\end{equation}
I3:
\begin{equation}
 \begin{aligned}
 \sin^{2}\theta_{13}&=\frac{1}{2}[\sin(\theta+\theta_{2})-\frac{1}{\sqrt{2}}\cos\theta\cos\theta {_2}]^2,\\
 \sin^{2}\theta_{23}&=\frac{1}{2}[\cos(\theta+\theta_{2})+\frac{1}{\sqrt{2}}\cos\theta\sin\theta {_2}]^2/(1-\sin^{2}\theta_{13}),\\
 \sin^{2}\theta_{12}&=\frac{1}{2}[\cos(\theta+\theta_{2})+\frac{1}{\sqrt{2}}\sin\theta\cos\theta {_2}]^2/(1-\sin^{2}\theta_{13});
\end{aligned}
\end{equation}
I4:
\begin{equation}
 \begin{aligned}
 \sin^{2}\theta_{13}&=\frac{1}{2}[\cos(\theta+\theta_{2})+\frac{1}{\sqrt{2}}\cos\theta\sin\theta {_2}]^2,\\
 \sin^{2}\theta_{23}&=\frac{3}{4}\cos^{2}(\theta-2\theta_{1})/(1-\sin^{2}\theta_{13}),\\
 \sin^{2}\theta_{12}&=\frac{1}{2}[\sin(\theta+\theta_{2})+\frac{1}{\sqrt{2}}\sin\theta\sin\theta {_2}]^2/(1-\sin^{2}\theta_{13}).
\end{aligned}
\end{equation}
The Dirac and Majorana CP phases are trivial for the above mixing patterns.

For the combination of Type V, we obtain the mixing pattern V1:
\begin{equation}
\label{eq:64}
U_{V1}=U^{+}_{e}(\theta_{2})\Omega_{\nu V}R_{12}(\theta)P_{\nu}S_{23}=\begin{array}{ccc}
         (V1_{1}
          & V1_{2} & V1_{3} )P^{\prime}_{\nu},
       \end{array}
\end{equation}
where
\begin{equation}
V1_{1}=\left(
                  \begin{array}{c}
                    \frac{e^{3i\pi/4}\sin\theta[(\sqrt{2}+\sqrt{3/2})\cos\theta_{2}-\sin\theta_{2}]+e^{i\pi/4}\cos\theta[(1-\sqrt{3})\cos\theta_{2}+(\sqrt{2}+\sqrt{6})\sin\theta_{2}]/2}{\sqrt{2(3+\sqrt{3})}} \\
             -\frac{1}{2}\cos(\theta-\theta_{1}-\theta_{2})-\frac{i}{2}\cos(\theta+\theta_{1}+\theta_{2})-\frac{\sqrt{2}}{4}\sin\theta_{2}[\cos(\theta-\theta_{1})+i\cos(\theta+\theta_{1})] \\
                   \frac{ 1}{4}(-12+6\sqrt{3})^{1/4}[(1+\sqrt{3})\cos\theta+i\sqrt{2}\sin\theta] \\
                  \end{array}
                \right),
\end{equation}
\begin{equation}
V1_{2}=\left(
\begin{array}{c}
\frac{\sqrt{3}}{2}\cos(\theta_{2}-2\theta_{1})\\
\frac{\sqrt{3}}{2}\sin(\theta_{2}-2\theta_{1})\\
-\frac{1}{2}
\end{array}
                \right),
\end{equation}
\begin{equation}
V1_{3}=\left(
                  \begin{array}{c}
                    \frac{1}{2}\cos(\theta-\theta_{1}-\theta_{2})-\frac{i}{2}\cos(\theta+\theta_{1}+\theta_{2})-\frac{\sqrt{2}}{4}\cos\theta_{2}[\sin(\theta-\theta_{1})+i\sin(\theta+\theta_{1})] \\
                    \frac{e^{-i\pi/4}\cos\theta[(\sqrt{2}+\sqrt{3/2})\sin\theta_{2}+\cos\theta_{2}]-e^{i\pi/4}\sin\theta[(\sqrt{2}+\sqrt{6})\cos\theta_{2}+(\sqrt{3}-1)\sin\theta_{2}]/2}{\sqrt{2(3+\sqrt{3})}} \\
                   \frac{ 1}{4}[(1-i)\sqrt{3-\sqrt{3}}\cos\theta+(1+i)(12+6\sqrt{3})^{1/4}\sin\theta]
                  \end{array}
                \right),
\end{equation}
and its permutation:
\begin{equation}
U_{V2}=S_{23}U_{V1}S_{13}.
\end{equation}
The mixing angles and CP invariants are listed as follows\\
V1:
\begin{equation}
\sin^{2}\theta_{13}=\frac{1}{16}[5 + \sqrt{ 3} \cos2\theta + \cos2\theta_{2} + \frac{7}{\sqrt{3}}\cos2\theta\cos2\theta_{2} - 2\sqrt{2}\sin2\theta_{2}-2 \sqrt{\frac{2}{3}}\cos2\theta\sin2\theta_{2}],\nonumber
\end{equation}
\begin{equation}
 \sin^{2}\theta_{23}=\frac{1}{16}[5 + \sqrt{ 3} \cos2\theta - \cos2\theta_{2} - \frac{7}{\sqrt{3}}\cos2\theta\cos2\theta_{2}
 + 2\sqrt{2}\sin2\theta_{2}+2\sqrt{\frac{2}{3}}\cos2\theta\sin2\theta_{2}]/(1-\sin^{2}\theta_{13}),\nonumber
 \end{equation}
\begin{equation}
 \sin^{2}\theta_{12}=\frac{3}{4}\cos^{2}(\theta_{2}-2\theta_{1})/(1-\sin^{2}\theta_{13}),\nonumber
 \end{equation}
\begin{equation}
 J_{cp}=\frac{3}{32}\sin2\theta\sin(4\theta_{1}-2\theta_{2}),
\end{equation}
\begin{equation}
J_{1}=(-1)^{k_{2}+1}\frac{\sqrt{3}}{64}\cos^{2}(\theta_{2}-2\theta_{1})[\sqrt{3}\cos2\theta(\cos2\theta_{2}-2\sqrt{2}\sin2\theta_{2}+5)
+2\sqrt{2}\sin2\theta_{2}-7\cos2\theta_{2}-3],\nonumber
\end{equation}
\begin{equation}
J_{2}=\frac{(-1)^{k_{1}}}{768} \sin2\theta(-27 \sqrt{2}-36 \sqrt{2} \cos2\theta_{2}+23\sqrt{2}\cos4\theta_{2}+72 \sin2\theta_{2}+52 \sin4 \theta_{2});\nonumber
\end{equation}

V2:
\begin{equation}
\sin^{2}\theta_{13}=\frac{1}{16}[5 +\cos2\theta_{2}- 2\sqrt{2}\sin2\theta_{2} + 2(1+\sqrt{3})\cos2\theta(-3-7\cos2\theta_{2} +2\sqrt{2}\sin2\theta_{2})],\nonumber
\end{equation}
\begin{equation}
 \sin^{2}\theta_{23}=\frac{\sqrt{12-6\sqrt{3}}}{16}[3+\sqrt{3}+(1+\sqrt{3})\cos2\theta]/(1-\sin^{2}\theta_{13}),\nonumber
 \end{equation}
\begin{equation}
 \sin^{2}\theta_{12}=\frac{3}{4}\cos^{2}(\theta_{2}-2\theta_{1})/(1-\sin^{2}\theta_{13}),\nonumber
 \end{equation}
\begin{equation}
 J_{cp}=\frac{3}{32}\sin2\theta\sin(4\theta_{1}-2\theta_{2}),
\end{equation}
\begin{equation}
J_{1}=(-1)^{k_{1}}\frac{\sqrt{3}}{64}\cos^{2}(\theta_{2}-2\theta_{1})[\sqrt{3}\cos2\theta(\cos2\theta_{2}-2\sqrt{2}\sin2\theta_{2}+5)
-2\sqrt{2}\sin2\theta_{2}+7\cos2\theta_{2}+3],\nonumber
\end{equation}
\begin{equation}
J_{2}=\frac{(-1)^{k_{2}+1}}{768} \sin2\theta(-27 \sqrt{2}-36 \sqrt{2} \cos2\theta_{2}+23\sqrt{2}\cos4\theta_{2}+72 \sin2\theta_{2}+52 \sin4 \theta_{2}),\nonumber
\end{equation}
where $J_{cp}$ denotes the Jarlskog invariant~\cite{47}, $k_{i}$ is the parameter introduced in Eq.~(\ref{eq:36}). The CP invariants are defined as
\begin{equation}
J_{cp}\equiv\mathrm{Im}[U_{11}U^{*}_{13}U^{*}_{31}U_{33}]=\frac{1}{8}\sin2\theta_{13}\sin2\theta_{23}\sin2\theta_{12}\cos\theta_{13}\sin\delta,
\end{equation}
 \begin{equation}
J_{1}\equiv\mathrm{Im}[(U_{11}^{*})^{2}U_{12}^{2}]=\sin^{2}\theta_{12}\cos^{2}\theta_{12}\cos^{4}\theta_{13}\sin\alpha,
\end{equation}
\begin{equation}
J_{2}\equiv\mathrm{Im}[(U_{11}^{*})^{2}U_{13}^{2}]=\sin^{2}\theta_{13}\cos^{2}\theta_{13}\cos^{2}\theta_{12}\sin\beta.
\end{equation}
Note that their magnitudes are invariant under the permutation of columns and rows of the mixing matrix.

\subsection{Numerical results of the viable mixing patterns}
\begin{figure}[tbp]
\centering % \begin{center}/\end{center} takes some additional vertical space
\includegraphics[width=0.48\textwidth]{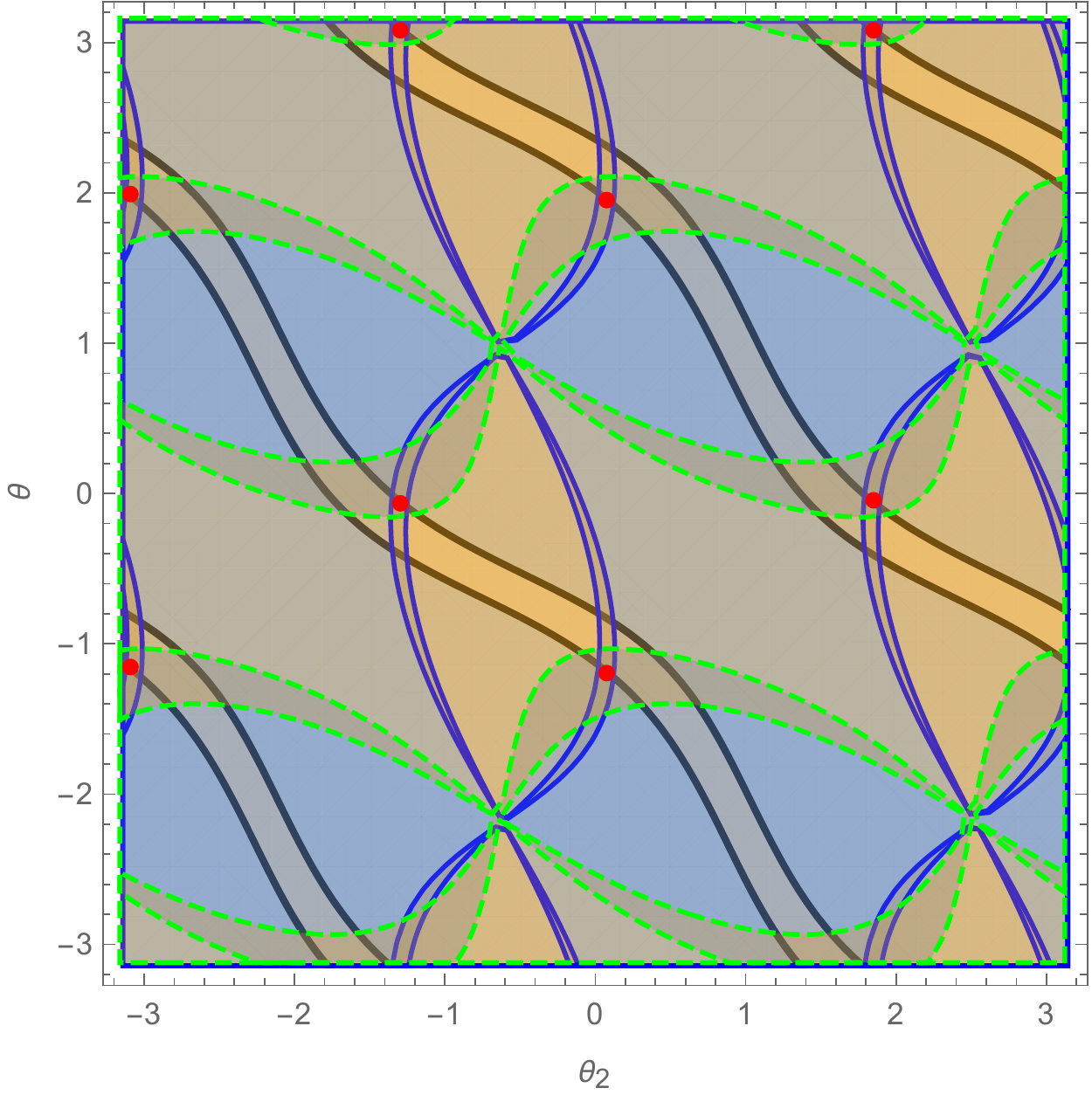}
\hfill
\includegraphics[width=0.48\textwidth]{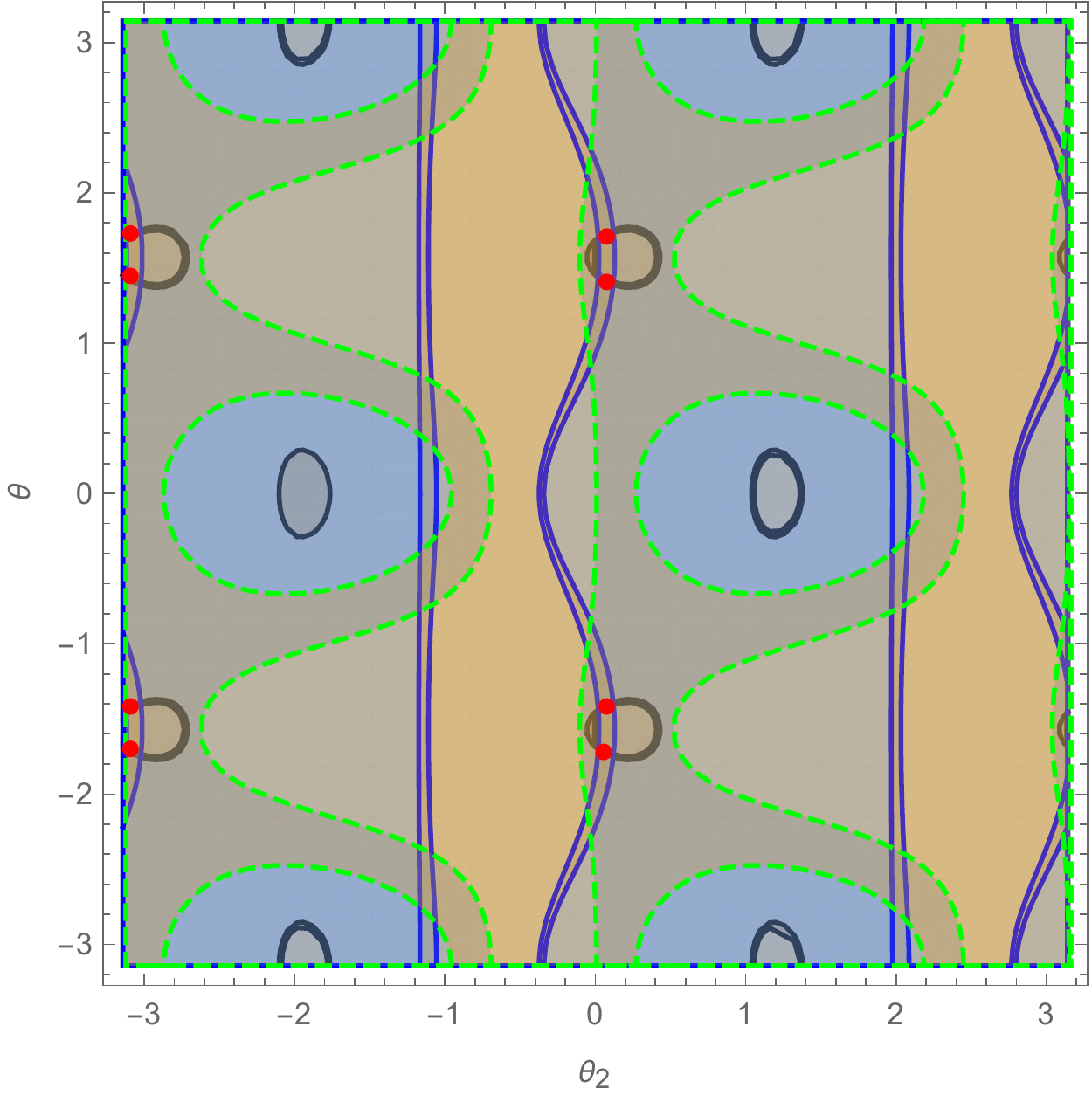}
\hfill
\caption{\label{Fig:1} Viable ranges of the parameters ($\theta_{2}, ~\theta$) for the mixing pattern I1 and VI at $3\sigma$ level. The left panel is for the pattern I1 and the right one for V1. Here the 3$\sigma$ ranges of mixing angles are taken from the fit data in Ref.~\cite{46}. In detail, $\sin^{2}\theta_{13}\in [0.01934,~0.02397]$ , $\sin^{2}\theta_{23}\in [0.385,~0.638]$,  $\sin^{2}\theta_{12}\in [0.271,~0.345]$. The viable ranges for $\theta_{12}$ at $3\sigma$ level are the strips with blue boundaries and those for $\theta_{23}$ are the areas with dashed green boundaries. The ranges of parameters for $\theta_{13}$ are tiny, i.e., reduced to  black curves. The intersections of these areas signed by red points show the final viable ranges for the mixing pattern. The parameter spaces for other viable mixing patterns are similar. So We don't shown them here.}
\end{figure}

As the quantitative prediction of the viable mixing patterns, we present the numerical results here.
First, we show the viable ranges of the parameters ($\theta_{2}, ~\theta$) for the mixing pattern I1 and VI at $3\sigma$ level in Fig.~\ref{Fig:1}. As is seen from the plots, the viable ranges of parameters for $\theta_{23}$ with green dashed boundary are considerable for both patterns. However, those for $\theta_{13}, \theta_{12}$ are tiny. So the final viable ranges of parameters are rather small. This observation holds for every viable patterns.

Second, we list best fit values of the leptonic mixing angles and CP phases of the viable mixing patterns in Table~\ref{tab:3}.
Let us give some comments on the results in the table:\\
(i). The best fit values of mixing angles $\theta_{ij}$ of every pattern in either normal or inverted ordering approximate those of the global fit dada in Ref.~\cite{46}.\\
(ii). When $1\sigma$ bounds are considered, the viable mixing patterns decrease. Because the Dirac CP violating phase in the mixing patterns of Type I is trivial, nether of these patterns could satisfy
the experimental constraints at $1\sigma$ level.
Furthermore, the mixing pattern V1 could not accommodate the mixing angles at $1\sigma $ level in either ordering. The only viable pattern at $1\sigma$ level is V2 in the case of inverted ordering. Its Dirac CP phase $\delta$ is around $\pm57^{\circ}$ or $\pm123^{\circ}$ which is in accordance with the result obtained in Ref.~\cite{41} and the global fit data~\cite{46}. For the Majorana phases, we obtain $\alpha\sim\pm90^{\circ}$, $\beta\sim\pm60^{\circ}$ or $\pm120^{\circ}$.
\\
(iii). We note a recent published paper~\cite{48} which does not appear to identify
$\Sigma(36\times3)$ as a viable candidate. Their results are based on the aforementioned combinations of residual symmetries in Case A where the $Z_{2}$ group hold in either neutrino or charged lepton sector. Our observation in this case is the same as theirs. However, in Case B which was not considered in their work, $Z_{2}$ group holds in both neutrino and charged lepton sectors. The results in this case shown in Table~\ref{tab:3}
reveals that the group $\Sigma(36\times3)$ with GCP could induce mixing patterns accommodating the experimental data.

\begin{table}
\caption{Numerical results for the mixing patterns of type-I, type-V. NO in the table refers to the normal ordering of neutrino masses and IO denotes the inverted ordering. }
\label{tab:3}       % Give a unique label
% For LaTeX tables use
\begin{tabular}{||c|c|c|c|c|c|c|c|c|c||}
\noalign{\smallskip}\hline
Patterns &~($\theta^{bf}$,~$\theta^{bf}_{2}$)~& ~~$\chi^{2}_{min}$~~&$~~\sin^{2}\theta_{13}$~~ & ~~$\sin^{2}\theta_{23}$~~ &~~ $\sin^{2}\theta_{12}$ ~& ~$|\sin\delta|$ ~& ~$|\sin\alpha|$& ~$|\sin\beta|$\\[0.5ex]\hline
\noalign{\smallskip}\noalign{\smallskip}\hline
I1(NO) & (-0.0171$\pi$,~0.587$\pi$),~(-0.3747$\pi$,~0.021$\pi$)~ &~~ 0.19~~ & 0.0216 & 0.451&0.304&0&0&0\\
I1(IO) & (-0.0067$\pi$,~0.575$\pi$),~(-0.385$\pi$,~0.033$\pi$)~ & 31.65 & 0.0221 & 0.474&0.333&0&0&0\\\hline
I2(NO) &  (-0.01$\pi$,~0.579$\pi$),~(-0.382$\pi$,~0.029$\pi)$~&17.22 & 0.0219 & 0.534&0.324&0&0&0\\
I2(IO) &  (-0.019$\pi$,~0.5896$\pi$),~(-0.373$\pi$,~0.0186$\pi)$ &2.77 & 0.0217 & 0.553&0.299&0&0&0\\\hline
I3(NO) & (-0.4897$\pi$,~0.424$\pi$),~(0.098$\pi$,~0.184$\pi)$~& 15.81 & 0.0221 & 0.512&0.337&0&0&0\\
I3(IO)& (-0.4599$\pi$,~0.402$\pi$),~(0.068$\pi$,~0.206$\pi)$& 0.09 & 0.0218 & 0.583&0.303&0&0&0\\\hline
I4(NO) &  (0.533$\pi$,~0.907$\pi$)),~(0.075$\pi$,~0.701$\pi)$& 0.25 & 0.0217 & 0.433&0.310&0&0&0\\
I4(IO) &  (0.5054$\pi$,~-0.072$\pi$),(0.329$\pi$,~2.134$\pi)$~& 25.92 & 0.0224 & 0.499&0.343&0&0&0\\\hline
V1(NO)&  ($\pm0.451\pi$,~0.022$\pi$)~~& 2.17 & 0.0217 & 0.406&0.307&0.849&0.9996&0.884\\
V1(IO) & ($\pm0.450\pi$,~0.024$\pi$)~~& 67.8 & 0.0218 & 0.406&0.311&0.863&0.9996&0.857\\\hline
V2(NO)&($\pm0.0493\pi$,~0.0232$\pi$)~~& 40.76 & 0.0217 & 0.594&0.309&0.858&0.9996&0.867\\
V2(IO) &($\pm0.0488\pi$,~0.0218$\pi$)~~& 0.11 & 0.0218 & 0.594&0.306&0.848&0.9996&0.887\\\hline
\noalign{\smallskip}
\end{tabular}
% Or use
\vspace*{0.5cm}  % with the correct table height
\end{table}

\section{Summary}
The GCP is an important approach to extract information on the CP phases of the lepton mixing pattern. We
employed the group $\Sigma(36\times3)$ with the GCP to predict lepton mixing patterns in a semidirect method. We first derived the general GCP which is compatible with the group $\Sigma(36\times3)$. Then we surveyed various combinations of Abelian residual flavor symmetries with the GCP , i.e., ($Z_{n(e)}$,~$Z_{2(\nu)}$,~$X_{\nu}$) and ($Z_{2(e)}$,~$X_{e}$,~$Z_{2(\nu)}$,~$X_{\nu}$).  We found two viable combinations (up to those equivalent) could accommodate the fit data of neutrinos at $3\sigma$ level. These combinations correspond to six mixing patterns among which four patterns predict trivial CP phases while two patterns give nontrivial ones.
In the nontrivial cases, the Dirac CP phase is around $\pm57^{\circ}$ or $\pm123^{\circ}$ which is in accordance with the result in the recent literature. When the experimental data at $1\sigma$ level is considered,
one of the two patterns predicting nontrivial CP phases is still viable in the inverted ordering.

We note that in the semidirect method the predictions for the six mixing parameters (three angles and three phases) depend on the free parameters. The physical reason for the best fit values of the free parameters is needed which is related to the open
question of the origin of the lepton mixing. As a compromise, we followed the line of the literature where the mixing patterns are partially determined
by the residual symmetries. So the fine-tuning of the free parameters is necessary in our work.  Even though, a dynamical model for the partial pattern is still important. However, for the flavor group $\Sigma(36\times3)$ which contains 108 elements and 14 irreducible representations, the model would be rather complicated in the technical aspect. So a dynamical model on the basis of $\Sigma(36\times3)$ with the GCP is out of the scope of this paper and we will consider it in the future work.

\acknowledgments
The author thank Zhi-zhong Xing, Yu-feng Li and Shun Zhou for warm hospitality at IHEP where the first version of the manuscript was finished. This work was supported by the National Natural Science Foundation of China under the Grant No. 11405101 and the foundation of Shaanxi University of Technology under the Grant No. SLGQD-13-10.
\\

\appendix{$\mathbf{Appendix}$}

\subsection{Equivalence of $X^{(i)}$}
The representations of $Z_{2}$ subgroups and the corresponding GCP are shown in Table~\ref{tab:2}. For the generator $b^{2}$, $X^{(1)}=E$, $X^{(2)}=\rho(b^{2})$. The unit matrix $E$ could be decomposed as $E=\Omega_{1}\Omega^{T}_{1}$, with
\begin{equation}
\Omega_{1}=\left(
             \begin{array}{ccc}
               0 & 1 & 0 \\
               \frac{1}{\sqrt{2}} & 0 & -\frac{1}{\sqrt{2}} \\
               \frac{1}{\sqrt{2}} & 0 & \frac{1}{\sqrt{2}} \\
             \end{array}
           \right).
\end{equation}
Note that $\Omega_{1}$ could also diagonalize $\rho(b^{2})$.
And the corresponding mixing matrix is $U_{\nu}=\Omega_{1}R_{12}(\theta)P_{\nu}$.
In contrast, $X^{(2)}=\Omega_{2}\Omega^{T}_{2}$, with
\begin{equation}
\Omega_{2}=\left(
             \begin{array}{ccc}
               0 & i & 0 \\
               \frac{i}{\sqrt{2}} & 0 & -\frac{1}{\sqrt{2}} \\
               \frac{i}{\sqrt{2}} & 0 & \frac{1}{\sqrt{2}} \\
             \end{array}
           \right)=\Omega_{1}\cdot diag(i, i, 1).
\end{equation}
And the mixing matrix is $U_{\nu}=\Omega_{2}R_{12}(\theta)P_{\nu}=\Omega_{1}R_{12}(\theta)P^{\prime}_{\nu}$, with
\begin{equation}
P^{\prime}_{\nu}=diag(i, i, 1)P_{\nu}.
\end{equation}
Therefore, the mixing pattern of $X^{(1)}$ is equivalent to that of $X^{(2)}$ up to a redefinition of the phase matrix $P_{\nu}$.\\
As for $X^{(3)}$ and $X^{(4)}$, we have $X^{(3)}=(X^{(4)})^{*}$. $X^{(3)}$ could be decomposed as $X^{(3)}=\Omega_{3}\Omega^{T}_{3}$, with
\begin{equation}
\Omega_{3}=\left(
             \begin{array}{ccc}
              -\sin\theta_{1} & \cos\theta_{1} & 0 \\
               \frac{\cos\theta_{1}}{\sqrt{2}} &  \frac{\sin\theta_{1}}{\sqrt{2}} & -\frac{1}{\sqrt{2}} \\
               \frac{\cos\theta_{1}}{\sqrt{2}} & \frac{\sin\theta_{1}}{\sqrt{2}} & \frac{1}{\sqrt{2}} \\
             \end{array}
           \right)\cdot diag(e^{i\pi/4}, e^{-i\pi/4}, 1 ),
\end{equation}
where $\theta_{1}=\arctan\sqrt{2}$. And $X^{(4)}$ could be decomposed as $X^{(4)}=\Omega_{4}\Omega^{T}_{4}$, with $\Omega_{4}=\Omega_{3}\cdot diag(i, i, 1)$. Thus the equivalence also holds for the mixing patterns of $X^{(3)}$ and $X^{(4)}$.

In the above discussion, we just consider the case of $Z^{b^{2}}_{2}$. For other subgroups $Z^{g_{\alpha}}_{2}$, the same observation still holds because of the similar transformation $\rho(g_{\alpha})=\Omega_{0}\rho(b^{2})\Omega^{+}_{0}$. In detail, the transformation $U_{\nu}(g_{\alpha})=\Omega_{0}U_{\nu}(b^{2})$ keeps the equivalence of $X^{(i)}$ in other cases.

\subsection{Equivalence of the mixing patterns from different residual symmetries}
The 3-dimensional representation of $\rho(g_{\alpha})$ with $g_{\alpha}\in Z_{2}$ and $X^{(i)}$ is shown in Table~\ref{tab:2}. On the base of the representation, we could obtain the matrix $\Omega_{i}$ which decomposes $X^{(i)}$, namely $X^{(i)}=\Omega_{i}\Omega^{T}_{i}$. With the matrices $\Omega_{i}$, the equivalence of the mixing patterns on the basis of combinations of Type I or Type V  could be examined. As an example, here we give transformations which relate the mixing patterns of Type V combinations.

For the combination $(Z^{ b^{2}}_{2(e)}$,~$E$, ~ $Z^{a^{2}b^{2}}_{2(\nu)}$, ~ $X^{(3)}=\rho(ab^{3}ab^{2}))$, the mixing matrix is obtained through the equation $U_{V}(\theta_{2},\theta)=U^{+}_{e}(\theta_{2})\Omega_{\nu V} R_{12}(\theta)P_{\nu}$ up to permutations of rows or columns, with
\begin{equation}
U_{e}=\Omega_{eI} R_{e 12}(\theta_{2})=
\left(
        \begin{array}{ccc}
          -\sin\theta_{2} & \cos\theta_{2} & 0 \\
          \frac{\sqrt{2}}{2}\cos\theta_{2} & \frac{\sqrt{2}}{2}\sin\theta_{2} & - \frac{\sqrt{2}}{2} \\
          \frac{\sqrt{2}}{2}\cos\theta_{2} & \frac{\sqrt{2}}{2}\sin\theta_{2} & \frac{\sqrt{2}}{2} \\
        \end{array}
      \right), ~~\Omega_{\nu V}=\left(
             \begin{array}{ccc}
               - \frac{e^{i\pi/4}\cos\theta_{1}}{\sqrt{2}} &  - \frac{e^{3i\pi/4}\sin\theta_{1}}{\sqrt{2}} & - \frac{\sqrt{2}}{2} \\
                - \frac{e^{i\pi/4}\cos\theta_{1}}{\sqrt{2}} & - \frac{e^{3i\pi/4}\sin\theta_{1}}{\sqrt{2}} &  \frac{\sqrt{2}}{2} \\
                e^{i\pi/4}\sin\theta_{1}& - e^{3i\pi/4}\cos\theta_{1}& 0 \\
             \end{array}
           \right),
\end{equation}
where $\theta_{1}=\frac{1}{2}\arctan\sqrt{2}$.
As for the combination $(Z^{ b^{2}}_{2(e)}$,~$E$, ~$Z^{cb^{2}}_{2(\nu)}$,~ $\rho(c^{2}bc^{2})$), we have
$U_{V^{\prime}}(\theta_{2},\theta)=U^{+}_{e}(\theta_{2})\Omega_{\nu V^{\prime}}R_{12}(\theta)P_{\nu}$,
where $\Omega_{\nu V^{\prime}}$ which decomposes $\rho(c^{2}bc^{2})$ is of the form
\begin{equation}
\Omega_{\nu V^{\prime}}=\left(
             \begin{array}{ccc}
               - e^{i\pi/4}\sin\theta_{1} & e^{i3\pi/4}\cos\theta_{1} & 0 \\
                \frac{e^{-i5\pi/12}\cos\theta_{1}}{\sqrt{2}} & \frac{e^{i\pi/12}\sin\theta_{1}}{\sqrt{2}} &  - \frac{\omega^{2}}{\sqrt{2}} \\
               \frac{e^{i11\pi/12}\cos\theta_{1}}{\sqrt{2}}& \frac{e^{-i7\pi/12}\sin\theta_{1}}{\sqrt{2}} & \frac{\omega}{\sqrt{2}}\\
             \end{array}
           \right).
\end{equation}
After a cumbersome derivation, we find the equivalent relation between them as follow
\begin{equation}
U_{V^{\prime}}(\theta_{2},\theta)=diag(1, -1, -i)U_{V}(-\theta_{2}+2\theta_{1},\pi-\theta)diag(1, -1, i).
\end{equation}
For the combination $(Z^{ b^{2}}_{2(e)}$,~$E$, ~$Z^{c^{2}b^{2}}_{2(\nu)}$,~ $\rho(cbc)$), we have
$U_{V^{\prime\prime}}(\theta_{2},\theta)=U^{+}_{e}(\theta_{2})\Omega_{\nu V^{\prime\prime}}R_{12}(\theta)P_{\nu}$,
with $\Omega_{\nu V^{\prime\prime}}$ of the form
\begin{equation}
\Omega_{\nu V^{\prime\prime}}=\left(
             \begin{array}{ccc}
                e^{i\pi/4}\sin\theta_{1} & -e^{i3\pi/4}\cos\theta_{1} & 0 \\
                -\frac{e^{i11\pi/12}\cos\theta_{1}}{\sqrt{2}} & -\frac{e^{-i7\pi/12}\sin\theta_{1}}{\sqrt{2}} &  - \frac{\omega}{\sqrt{2}} \\
               -\frac{e^{-i5\pi/12}\cos\theta_{1}}{\sqrt{2}}& -\frac{e^{i\pi/12}\sin\theta_{1}}{\sqrt{2}} & \frac{\omega^{2}}{\sqrt{2}}\\
             \end{array}
           \right).
\end{equation}
And the equivalent relation between the mixing matrices is
$U_{V^{\prime\prime}}(\theta_{2},\theta)=diag(-1, -1, 1)\cdot U_{V^{\prime}}(\theta_{2},\theta)$.
Therefore, on the basis of above discussions, we find that different residual symmetries could correspond to the same mixing pattern through the redefinitions of phase matrix and free parameters.

\end{document}